\documentclass{nature}
\usepackage{pdfpages}
\usepackage[]{algorithm2e}
\usepackage{graphicx}
\usepackage[round]{natbib}
\usepackage{fancyhdr} 
\makeatletter
\let\saved@includegraphics\includegraphics
\AtBeginDocument{\let\includegraphics\saved@includegraphics}
\renewenvironment*{figure}{\@float{figure}}{\end@float}
\makeatother

\title{Application of electromagnetic centroids to colocalization of fluorescing objects in tissue sections}

\author{Renata Rycht\'{a}rikov\'{a}$^{1,*}$, Georg Steiner$^{2}$, Gero Kramer$^{3}$, Michael B. Fischer$^{4}$, Dalibor \v{S}tys$^{1}$}

\begin{document}

\maketitle

\begin{affiliations}
 \item University of South Bohemia in \v{C}esk\'{e} Bud\v{e}jovice, Faculty of Fisheries and Protection of Waters, South Bohemian Research Center of Aquaculture and Biodiversity of Hydrocenoses, Kompetenzzentrum MechanoBiologie in Regenerativer Medizin, Institute of Complex Systems, Z\'{a}mek 136, 373 33 Nov\'{e} Hrady, Czech Republic
 \item TissueGnostics GmbH, Taborstrasse 10/2/8, 1020 Vienna, Austria
 \item Medical University of Vienna, Department of Urology, W\"{a}hringer G\"{u}rtel 18-20, 1090 Vienna, Austria
 \item Danube University Krems, Department of Health Sciences and Biomedicine, Dr.-Karl-Dorrek-Strasse 30, 3500 Krems an der Donau, Austria
 \newline $^{*}$ \textbf{Corresponding author:} Email: rrychtarikova@frov.jcu.cz; Telephone: +420-38 777 3833
\end{affiliations}

\begin{abstract}
Light microscopy as well as image acquisition and processing suffer from physical and technical prejudices which preclude a correct interpretation of biological observations which can be reflected in, e.g., medical and pharmacological praxis. Using the examples of a diffracting microbead and fluorescently labelled tissue, this article clarifies some ignored aspects of image build-up in the light microscope and introduce algorithms for maximal extraction of information from the 3D microscopic experiments.
We provided a correct set-up of the microscope and we sought a voxel (3D pixel) called an electromagnetic centroid which localizes the information about the object. In diffraction imaging and light emission, this voxel shows a minimal intensity change in two consecutive optical cuts. This approach further enabled us to identify z-stack of a DAPI-stained tissue section where at least one object of a relevant fluorescent marker was in focus. The spatial corrections (overlaps) of the DAPI-labelled region with in-focus autofluorescent regions then enabled us to co-localize these three regions in the optimal way when considering physical laws and information theory. We demonstrate that superresolution down to the Nobelish level can be obtained from commonplace widefield bright-field and fluorescence microscopy and bring new perspectives on co-localization in fluorescent microscopy.
\end{abstract}

\noindent \textbf{Keywords:} point divergence gain, electromagnetic centroid, superresolution microscopy, 3D fluorescence colocalization, microscope construction, bright-field light microscopy

\section{Introduction}

The understanding of the physico-chemical basis of the intracellular processes requires determination of local concentrations of cell chemical constituents. For that, the optical microscopy is the irreplaceable method. However, in most cases the biological observations are interpreted from a projection which was captured at one ``compromise" or ``optimal" focal plane. Most analyses of co-localisation of fluorescent labels are impaired by misalignment caused by different positions of the focuses of individual colour channels or fluorophores. Moreover, the spatial resolution of the optical microscope, or, more precisely, the facility to solve the inverse problem of determination of the spatial location of the object which gives rise to the observed signal and its individual electromagnetic nature, is limited.

Here we analyze deviations of the intensity profile of the electromagnetic field (light) due to its interactions with matter and following modification along the microscope optical  path~\citep{Thorn2016}. In interactions with cells, we can mainly observe: (i) the diffraction of light on organelles, macromolecular complexes and other intracellular structures and (ii) the fluorescence emission of autofluorescent or fluorescently labelled objects. 


The diffraction of light is a process whose complete explanation is complicated and, for the majority of real cases, even practically impossible~\citep{Mie1909}. A common view of light passing behind a scattering object is a picture of a dark ``cone" contracting as the light waves enfold the gap. When the object is circular enough, a bright spot on axis, called the Arago spot~\citep{Fresnel1868}, arises. Biological objects consist of dense matter made of proteins, nucleic acids, lipids, and other molecules. Diffracting elements inside the living cell differ in their refractivity index from that of their surroundings and are internally inhomogeneous, as well. Light scattering can characterize the spectral properties of intracellular objects properly. All we obtain after the light passes through the biological sample and the microscope is `information' from the microscopy experiment. About the origin of this kind of information, we have only limited knowledge.  

Fluorescent microscopy of a living cell~\citep{Thorn2016} has numerous advantages over the diffraction observation, namely a very small light source that is a single chemical bond. This enables experimentalists to explore and exploit different limits of the modification of light along the light path of the microscope separately. The breaking of the long-established resolution limit~\citep{Abbe1874} of the light microscope was awarded the Nobel Prize~\citep{nobelprize2016}. These so-called superresolution methods~\citep{bohm2016,hell1994,hell1995,hell2005,betzig2006} can be explained using the electromagnetic field theory~\citep{Maxwell}. This theory preceded quantum mechanics with its additional uncertainty limit given by the Heisenberg principle~\citep{Heisenberg1927}, which describes the behaviour of a single photon and is seemingly contradicted by superresolution. The Maxwell theory, in contrast, describes well the behaviour of an ensemble of photons. When the maximum of this ensemble is sought, superresolution is not in conflict with contemporary theories of light.   

As we have shown~\citep{Rychtarikovaetal2017,Rychtarikovaetal2017b}, the information reaching the image sensor of a digital camera can be scrutinized down to the level of a its single element, i.e. of a single pixel in a digital image. In other words, the information limit is given by the size of an element which is theoretically projected on a single pixel of the digital camera. The size of this elementary information may be experimentally determined and can have an area of a few squared nanometres. As discussed above, no principles of quantum mechanics are broken if we determine the location of the distribution profile for a sufficiently large ensemble of photons. In such a distribution it is then possible to seek a maximum, a median, skewness, etc. 

The location of the intensity extreme does not determine the position of the object, which causes the change in the electromagnetic field profile. For instance, in diffractive imaging the darkest and smallest point is located outside the object at a position governed by the diffraction process. The experimentally determined light intensity maximum or minimum is found, in the proper definition (for all types of imaging), as a centroid of the outcome of the electromagnetic process. Later in the text, this point is called the electromagnetic centroid and its existence is in compliance with the Extended Nijboer-Zernike Theory, e.g.,~\citep{BraatJannsen2015}.

The analysis of information from the digital image also includes a description of all non-idealities of the optical path. In a colour digital camera, each camera channel catches different information~\citep{Rychtarikovaetal2017,Stysetal2016,Cisaretal2016}. This difference is caused by the differences in chemical composition of the observed object that gives rise to the signal. The exact way of the transfer of this difference by the microscope to the camera chip needs to be examined in detail. For instance, contemporary apochromatic objectives utilize combinations of lenses to project all colours at the same place. This may be achieved only in idealised samples and with a finite precision. When minute details of the microscopic image are interpreted, the apochromaticity cannot be expected and the properties of the light path should be experimentally examined for each lens separately. 

As we have summarized~\citep{Rychtarikovaetal2017b}, the microscopic observation should answer,

\begin{enumerate}
\item where the object giving rise to the response is located,
\item what the shape of the object is, and
\item what the spectral characteristics of the object are.
\end{enumerate}

We have processed microscopic z-stacks of (1) a single microparticle and (2) a section of fluorescently labelled prostate cancer tissue. Based on the physico-technical aspects mentioned above, with the precision of a single voxel, we have systematically determined all electromagnetic centroids of the diffracting microbead as information centroids in 3D space. We have extended this approach to widefield fluorescence micrographs of a tissue section in which the electromagnetic centroids and the projected positions of the light emitting object are at the same place. With the respect to the construction of the light microscope, we developed specific algorithms based on the R\'{e}nyi information theory which respects the multifractality of the observed object. This complex image multifractality arises from the transfer of the information from an observed specimen through a microscope optical path up to a digital camera chip. These algorithms then enable us to detect different modes of binding of fluorophores, fluorophore densities and intensities of the emission in the raw image data (i.e., data untreated by camera software). These analyses demonstrate the power of the approach.

In this article, we also demonstrate that the specificity of the fluorescent labelling can be improved. We demonstrate the enormous increase of the intelligibility of the primary, raw, fluorescence image data. Thus, cell microscopy is dominated by an idea that if, in an image, there are two colours (emitted wavelengths) projected at the same point, then these colours are co-localized. However, when we analyze the image signal in 3D, we can determine that each colour (wavelength) is projected onto a different point of space. The conclusions on co-localization can be made only after alignment of the focal planes of the colours and, in this paper, we provide an objective, information-entropic, tool for such a correction and compare it with the standard microscope's auto-focusing.

\section{Materials and methods} \label{sec:methods}

\subsection{Microscopy and image processing of a diffracting microparticle}
A 2-$\mu$m latex particle placed on a carbon layer on a electron microscopy copper grid covered by amorphous carbon (obtained at the Institute of Parasitology AS CR, \v{C}esk\'{e} Bud\v{e}jovice, CZ) was scanned under a inverted light transmission microscope~\citep{Rychtarikovaetal2017,Rychtarikovaetal2015} (Institute of Complex Systems, Nov\'{e} Hrady, CZ) equipped by a 12-bit-per-channel colour Kodak KAI-16000 digital camera with a chip of 4872$\times$3248 resolution (Camera Offset 200, Camera Gain 383, Camera Exposure 2950 ms) and home-made control software. A Nikon objective (60$\times$/0.8, $\infty$/0.17, WD 0.3), which gives the resulting size of an image pixel as 46$\times$46 nm$^2$ per each camera channel, was used. The sample was illuminated by two Luminus 360 LEDs charged by the current of 4500 mA. The pngparser.exe software~\citep{Rychtarikovaetal2017} classified the micrographs to give a z-stack of 258 images of the average z-step of 152 nm.

The procedure for finding the electromagnetic centroids in each image colour channel is schematically described in Fig.~1. In each raw file (primary image data) of an optical cut, vignetting of the microscope optical path was suppressed (line 1 in Fig.~1) by calibration of each camera pixel using a set of calibration in-focus images of 20-$\mu$m NDL-10S-4 - Step Variable ND metallic filters (OD: 0.1--4.0; ThorLabs) with their relevant transparent spectra acquired using an Ocean Optics USB 4000 VIS-NIR-ES spectrophotometer whose fiber's top was positioned at the place of the microscope objective. The image calibration and correction themselves were performed using a VerCa$\_$cmd 0.1.6 software (Institute of Complex Systems, Nov\'{e} Hrady, CZ). The basic principle of the correction of image intensities by the VerCa$\_$cmd software is in detail described as Supplementary Material in Arxiv:1908.03696 (since 14 Aug 2019). The corrected micrographs of the microbead showed the 14-bit-per-channel depth.

First of all, in each optical cut, the distracting background was manually separated from the microbead's image -- point spread function (PSF). After that, each pair of the corrected consecutive micrographs was subtracted in order to localize the pixels of unchanged intensities as zero points. These zero points were thresholded to give binary images that were multiplied by the original micrographs. After separation of each colour (red, green 1, green 2, and blue) channel, the electromagnetic centroids were detected as local intensity extrema.

\subsection{Microscopy and image processing of prostate cancer tissue}
A section of prostate cancer tissue was fixed and imunnofluorescently labelled by 4',6-diamidine-2'-phenylindole (DAPI). 

The sample was then scanned with the z-step of 100 nm using a TissueFaxs-PLUS-Confocal fluorescent microscope (TissueGnostics GmbH, Vienna, AT) based on an AxioImager Z2 and  TissueFaxs 5.2 instrument control software (TissueGnostics, Austria). 
The light source was a SPECTRA X LED (Lumencor, Beaverton, USA), the confocal device was a spinning disc device X-Light V2 (CrestOptics, Roma, Italy). The microscope was further equipped by a 16-bit grayscale camera Orca flash 4.0 (Hamamatsu, Hamamatsu City, Japan) with a chip of 1560$\times$1960 resolution. A 100$\times$ oil objective gave the resulting image pixel of the size of 328$\times$328 nm$^2$. The full z-stacks (DAPI in the blue channel and autofluorescence in red and green channel) contained 81 14-bit images. The real z-position of each image was read out from the name of the image.

The monochrome z-stacks of the fluorescing prostate cancer tissue (Fig.~2) were processed in a way to find the electromagnetic centroids in uncorrected images. For this, we used computation based on the parametrized R\'{e}nyi information entropy as described in~\citep{pdg}. This approach utilizes a novel variable -- point divergence gain -- which, due to systematic changes in values of the multifractality-respecting parameter $\alpha$ enables us to find and subtract intensities of close values, not only intensities of the same value, placed above each other and, in this way, suppress the image intensity noise. In addition, with the usage of the additive variables of the point divergence gain, this approach allows us to specifically describe each series image using $\alpha$-dependent spectra.

The computation began with the selection of an in-focus region via computation of the two additive variables of the point divergence gain -- point divergence gain entropy
\begin{equation}
I_\alpha = \frac{1}{\mid 1-\alpha \mid}\sum_{i = 1}^{n} \mid \log_2 \frac{\sum_{i=1}^{j} p_{k/l}^\alpha}{\sum_{i=1}^{j} p_{i}^\alpha}\mid
\label{Eq1}
\end{equation}
and point divergence gain entropy density
\begin{equation}
P_\alpha = \frac{1}{\mid 1-\alpha \mid}\sum_{i = 1}^{m} \mid \log_2 \frac{\sum_{i=1}^{j} p_{k/l}^\alpha}{\sum_{i=1}^{j} p_{i}^\alpha} \mid
\label{Eq2}
\end{equation}
for the set of R\'{e}nyi coefficient $\alpha$ = \{0.1, 0.3, 0.5, 0.7, 0.99, 1.3, 1.5, 1.7, 2.0, 2.5, 3.0, 3.5, 4.0, 5.0, 6.0\}. In Eqs.~1--2, the $p_{k/l}$ is a probability of the occurrence of the pixel of intensity $k$ after exchanging for the pixel of intensity $l$ at the same position in the following image and the $p_i$ is a probability of the occurrence of each intensity in the image with the intensity $k$. Variable $i$ = \{1,2, ..., $k$, ..., $l$, ..., $j$\} is the label of a bin in the intensity histogram; the $n$ corresponds to the number of pixels in the image and the $m$ corresponds to the number of pixels of unique values of the point divergence gain (see Eq.~3). This type of calculation was performed using a Image Info Extractor Professional software (IIEP; Institute of Complex Systems USB, Czech Republic). For all three image series, all resulted $\alpha$-dependent spectra of the $I_\alpha$ and $P_\alpha$ gave a matrix of the size of 81$\times$30 where each matrix column unambiguously describe each series image. This matrix could be therefore clustered by k-means++ algorithm (with squared Euclidian metrics~\citep{Arthur}) into two (in-focus and out-of-focus) groups. The middle -- in-focus -- image subseries was separated. We found the in-focus range from img. 26 to 77 for red autofluorescence and DAPI and from img. 1 to 66 for green autofluorescence.

The electromagnetic centroids themselves were being searched in the in-focus series by the direct computation (set-up of the IIEP software: Ignore black pixels) of the above-mentioned point divergence gains~\citep{pdg} 

\begin{equation}
\Omega_{\alpha, k/l} = \frac{1}{1-\alpha}\log_2 \frac{\sum_{i=1}^{j} p_{k/l}^\alpha}{\sum_{i=1}^{j} p_{i}^\alpha},
\label{Eq3}
\end{equation}
which are parts (a summand) of Eqs.~Eq1--2 and define information changes after an exchange of one pixel for another one at the same position in the next image. For each pair of the in-focus images, the positions of all zero $\Omega_{\alpha, k/l}$ at $\alpha = 6$ for red and green autofluorescence and at $\alpha = 7$ for DAPI were found to give a binary mask for extraction of relevant points of the point spread function (PSF; lines 1--14 in algorithm~in Supplementary Material 2). This was followed by selection of objects on the basis of either Otsu's thresholding or by a combination of the image morphological operations (lines 15--16 in algorithm~in Supplementary Material 2). In this way, we modelled a theoretical interlayer of the bead's PSF. All interlayers were stacked into a 3D matrix.

Since the in-focus series of red autofluorescence and DAPI had the same number of images and red and green autofluorescence are expected to mark the same (similar) intracellular structures, the in-focus images of fluorescent labels were co-localized via overlapping of the relevant images of the red autofluorescence and DAPI followed by the alignment of the information foci (the image with the lowest value of the $I_{0.99}$ in the green channel) of green autofluorescence with red autofluorescence (in details in Results: Fluorescence microscopy of tissue section).

\begin{algorithm} \label{Alg2}
\IncMargin{1em}
\LinesNumbered
\KwIn{a focused region of $N_f$ grayscale images (of the size of $x\times y$ pixels) from fluorescent microscope\\ \qquad \quad
$r$-1 matrices of the $\Omega_{\alpha, k/l}$ (in .mat files)\\ \qquad \quad
\textbf{M} as a zero $x\times y\times N_f$ matrix
}
\KwOut{\textbf{distM} as a 3D matrix of large fluorescently labelled objects\\ \qquad \quad
\textbf{otsuM} as a 3D matrix of strongly fluorescing objects}
\BlankLine
\BlankLine
\For{$i = 2$ \KwTo $N_f$}{
\textbf{PDG1} = readMat($i$);\\
\textbf{PDG2} = readMat($i$-1);\\
\qquad\emph{\% read two consecutive --(i)$^{th}$ and (i-1)$^{th}$-- .mat files containing the $\Omega_{\alpha, k/l}$}\\
\textbf{zPDG1} = (\textbf{PDG1} == 0);\\
\textbf{zPDG2} = (\textbf{PDG2} == 0);\\
\qquad\emph{\% in two consecutive images, threshold zero $\Omega_{\alpha, k/l}$}\\
\textbf{joinI} = (\textbf{zPDG1} + \textbf{zPDG2});\\
\qquad\emph{\% sum the two logical images with positions of $\Omega_{\alpha, k/l}$ = 0}\\
\textbf{BI} = uint(\textbf{joinI} $>$ 0);
\qquad\emph{\% binarize the joint image for all non-zero values}\\
\textbf{img} = readIm($i$);
\qquad\qquad\emph{\% read the (i)$^{th}$ grayscale image}\\
\textbf{M}($i$-1) = \textbf{img} .* \textbf{BI};\\
\qquad\emph{\% save the intensities at the position of zero $\Omega_{\alpha, k/l}$ into the (i-1)$^{th}$ layer of the matrix \textbf{M}}\\
}
\BlankLine
\BlankLine
\textbf{distM} = remSmallObj(\textbf{M}) $||$ \textbf{otsuM} = otsu(\textbf{M}); \\ \qquad \emph{\% remove small irrelevant objects or undesirable low intensities from the matrix \textbf{\mbox{M}}}\\
\BlankLine
\BlankLine
\caption{Algorithm for 3D reconstructions using zero $\Omega_{\alpha,k/l}$}
\end{algorithm}

\subsection{Data visualization}
\label{processing}
Processing of primary microscopic data was similar to the procedure published previously in~\citep{Rychtarikovaetal2017}. Figures in the article were visualized using the LIL conversion into 8 bits and plotted using Matlab$^{\textregistered}$ 2016b (Mathworks, USA) software.

\section{Results} \label{sec:results}
\subsection{Electromagnetic centroid of a diffractive object}
\label{subsec:centroid}

Fig.~3\textbf{a,c} shows optical cuts through a 3D image of a 2-$\mu$m latex bead from a widefield transmission microscope. Fig.~3\textbf{a} depicts the relevant xz- and yz-planes, whereas Fig.~3\textbf{c} shows z-planes, i.e., real intensity images. In each colour (red, green, and blue) channel, the intensity distributions (PSFs) which are a consequence of the interaction of light with the object have different shapes and positions. It is a combination of real-life diffraction behaviour, the mixed response of different wavelengths, which are summed in each colour channel, and of all non-idealities of the microscope optics and camera electronics. In other words, the presented intensities are combinations of an interaction of the photons with the sample followed by a transformation of the intensities during passage of the photons through the microscope optical system and by the interaction of the photons with camera sensors. 

Independently of the size of the observed objected, in the green and blue channels, we experimentally determined two voxel-sized objects along/close to the z-axis of diffracting object's (bead's) PSF which, in these spectrally calibrated and corrected images, show the same intensity for two consecutive images and are intensity extremes (Fig.~3\textbf{b}): (1) The dark spot is an outcome of the shading, non-idealities of light behaviour and destructive light diffraction and represents the densest information on the object. We called this spot the \textit{electromagnetic centroid}. This dark spot is followed by (2) the brightest spot, which is a result of positive light interferences and is surrounded by dark intensities. This can be a real-life manifestation of the Arago spot~\citep{Fresnel1868}. For the red channel, the darkest point is localized out of the z-axis due to the optics aberrations.


\subsection{Fluorescence microscopy of tissue section}
\label{subsec:cells}

Fig.~4 shows a micrograph of a fluorescing tissue visualised using a standard pseudocolour approach. The observed intensities are a result of the light emission of fluorophores localized at particular positions in the sample. A correct way how to localize the positions of the fluorophores with the precision of one voxel is the application of information analysis of the given z-stack using the point divergence gain~\citep{Rychtarikovaetal2017,Rychtarikovaetal2015} which defines the electromagnetic centroids as the brightest points of the unchanged information/intensity between two consecutive images. The relative variables point information gain entropy and point information gain entropy density, based on the characterization of the image multifractality, contain information about positions of electromagnetic centroids in optical cuts and, thus, can describe information changes in consecutive images. In these aspects, these two variables, in the form of $\alpha$-dependent spectra, are a suitable tool for image data clustering such as searching for an in-focus range. Moreover, the approach of multifractality is strong enough for suppression of image intensity noise and, thus, for finding the electromagnetic centroids in images of a spectrally uncalibrated microscope optical path.

The z-stacks were acquired in a way that for each fluorophore the microscope found a focus automatically and the microscope was then set-up to scan 40 images symmetrically below and above this focus. In these ranges of 81 images, the information-entropic analysis based on the point information gain entropy and point information gain entropy density (see Materials and Methods: Microscopy and image processing of prostate cancer tissue) identified in-focus images where at least one fluorophore should be in the focus. Similar to the standard autofocusing of the microscope, this method will determine the so-called information focus, which is an image where the majority of the fluorophores are localized. These positions correspond to the image of the minimal (non-moving objects autofluorescing in red and green) or maximal (DAPI-labelled objects) value of the $I_{0.99}$ (a sum of all point divergence gains in an image). Generally, in the case of non-moving (stable in time) objects, the number of zero $\Omega_{0.99, k/l}$ gradually decrease up to the focus (read in detail below). In contrast, in the information focus, moving objects change their spectral properties along the z-axis substantially more than they do out of the focus which increases the average value of the $\Omega_{0.99, k/l}$ in the image. The unique value of $\alpha$ = 0.99 (approximation of the R\'{e}nyi entropy to the Shannon entropy) was chosen, since the intensity histogram of the given images exhibit normal-like distributions.
 
The real positions and ranges of the z-stacks for the standard, Fourier-transform based, microscope's autofocusing and those obtained via the information-entropic analysis are compared in~Fig.~5\textbf{a} and \textbf{b}. The focal region of the green autofluorescence was broader than the focal regions of the red autofluorescence and DAPI. This can be attributed to the broader green part of the light spectrum that is projected along the broader optical path due to chromatic aberration. Both DAPI foci correspond to the same image (img. 41). The standard autofoci of the red and green autofluorescence are shifted about 434 nm and 1,000 nm, respectively, above the DAPI focus. Whereas the information focus of the red autofluorescence lies about 1,100 nm (about 11 images) above its standard autofocus, the green autofluorescence has its information focus about 200 nm (about 2 images) lower than its standard focus. After the information-entropic clustering, the ranges of all z-stacks were narrowed so that they were limited by the images with the points of the highest emission and their images were asymmetrically distributed around their information foci.

The next step in the analysis was to find points of zero point divergence gain, i.e. points whose intensity minimally changes over two z-levels, in a series of images~\citep{Rychtarikovaetal2017,Rychtarikovaetal2015}. This analysis spatially localizes all in-focus objects that are giving rise to the fluorescence emission. The amount of points $\Omega_{6.0}$ = 0 at each level is relatively low and the information complexity of the image is reduced significantly (Fig.~6).

Specific fluorescent labelling is often a basis of interpretation of biological data at the cell or tissue level. In case that labelling is not specific enough, i.e. fluorescing points are distributed in the whole image, we can define rules according to which the examined points differ from the others. This is explained in Figs.~7--8. In Fig.~7\textbf{c--d}, \textit{left}, Fig.~8\textbf{a} and Video S1, where on the basis of existence of a voxel with in-focus emission in the close proximity of another voxel with in-focus emission (dense areas), we identified nuclei. This statement is further strengthened by the fact that we deal with ellipsoidal objects labelled by DAPI. However, we identified another class of DAPI-labelled objects which dominate in Fig.~7\textbf{c--d}, \textit{right}, Fig.~8\textbf{b} and Video S2. These objects were selected on the basis of the different intensities of the DAPI emission in the in-focus voxels, i.e. there was specific binding but it was more scattered. However, as seen in Fig.~7\textbf{a--b}, any simple rule for the selection of points is often not available (or necessary) and the image has to (or can be) analyzed from the complete dataset. In this case, the 3D model was reconstructed from micrographs in which the pixels of unchanged intensity occurred only very sparsely. This indicates that we deal with autofluorescence. Autofluorescence can be typically better localized than inserted fluorescence.

DAPI with the autofluorescence in the red and green regions of the light spectrum were co-localized by the procedure which is visualized in Fig.~5\textbf{a}. The whole z-stack of images of red autofluorescence was, in agreement with the focus determination calculation, shifted down by 434 nm to overlap the z-stack with DAPI-labelled objects. The green fluorescing z-stack was shifted by 300 nm up to reach the red focus. This treatment gave the resulting RGB stack where the course of the $I_{0.99}$ for green autofluorescence copies this course for red autofluorescence. The focus of DAPI was brought even closer to the others (Fig.~5\textbf{b}) with the shift of 1,200 nm.

With the example of autofluorescence in red and green regions, we shall further explain technical aspects of the focal plane location and co-localization. Object O1 (Fig.~8\textbf{a} and Video S1) is autofluorescing in green and red in images 1--50 and 40--51, respectively. This phenomena can be only explained by the different projection of each emitted light wavelength along the light path of the microscope. In many other objects,  e.g. O2 in Fig.~8\textbf{b} and Video S2, both colour channels co-localize at all levels where the green and red autofluoresce is in focus. We probably deal with autofluorescing filaments which span the whole sample (cf. Fig.~7\textbf{a,b}).

Regarding DAPI, its co-localization with red and green autofluorescence in nuclei is marginal and occurs only at the surface of the nuclei (Fig.~8\textbf{a} and Video S1). In contrast, inside the sparsely DAPI-labelled objects in Fig.~8\textbf{b} and Video S2,  a significant co-localization of this type was found. This further confirms the above-mentioned claim that these regions are not nuclei but other organelles which are spanned by the autofluorescing filaments. Without the correct 3D analysis, these labelled cell compartments would be mixed up with ``true" nuclei.

\section{Discussion}
\label{sec:discussion}

This paper describes how to objectively localize unchanged information between two images which are shifted along the optical axis (optical cuts) in light microscopy in order to acquire superresolution micrographs. Using this procedure it is possible to find the most localized information about the position of an object, so-called electromagnetic centroid. General technical requirements for construction of a (superresolution) microscope are summarized in, e.g.,~\citep{McNamara,Godin}. In this discussion, we would like to stress those aspects that are important for finding the electromagnetic centroids and co-localization in biological imaging using widefield light microscopy:
\begin{enumerate}
\item The comparison of the optical cuts is limited by the analog-to-digital (AD) conversion. The primary data provided by the AD converters in standard digital cameras are of a 12-, 14- or 16-bit intensity depth. However, experimenters often visually analyse images in their 8-bit representations which are heavily distorted by a standard conversion from the vice-bit data~\citep{Stysetal2016}. In this paper, we analyse original vice-bit datasets. 

\item The theoretical size of the observed object is given by a size of the pixel or voxel. In other words, the object's discriminability as defined in~\citep{Rychtarikovaetal2017b} is determined experimentally by the size of the camera pixel, by the number of the camera pixels per chip area, and by the step of the microscope mechanics along the optical axis. 

\item Intensity calibration of a purchased camera chip is insufficient. Moreover, there always exists a concern that the camera sends out a signal modified by a software whose functions are hidden to the experimenter. Let us note that the responses of the chip elements to the exposure time and light intensity is not linear.

\item Not only the camera chip but the whole optical path needs to be calibrated in order to compensate its non-idealities.

\item At longer acquisition times, moving objects like living cells give a false negative signal. 

\item In the examples presented here, i.e., in the samples that were observed at high light intensities 
, the key limit that prevents finding the intensity maximum or minimum is not most probably the quantum noise~\citep{Mizushima1988}. The distortions of the signal were generated along the optical path and in the instrument electronics. 
Precise characteristics of the noise are unknown. Most generally, the noise may be assumed to have a multifractal character and is discretized in space and in time. The calculation of point divergence gain~\citep{Rychtarikovaetal2017,Rychtarikovaetal2015} enables us to analyse the whole distribution of noise distortions. The points (pixels) are grouped according to different multifractal properties. In the diffractive imaging, we used this method~\citep{Rychtarikovaetal2017,Rychtarikovaetal2015} to obtain the elementary information contributions in z-stacks of images of living cells. 
\end{enumerate}

A 2-$\mu$m microparticle has a size similar to that of mitochondria and other oval organelles. The size of this bead in the reported experiment ranges between 5 light waves (shortwave blue, 400 nm) and less than 3 light waves (far red, 720 nm). In other words, this size attributes the bead both macroscopic behaviour and nanoscopic behaviour at which the quantum dot effect is observed. 
In the macroscopic description, we interpret our observation as a constructive interference arising behind the dark area of diffraction.

The image analysis of a single microparticle/organelle was performed using two basic approaches of the signal analysis~\citep{Urbanetal2015}: the resolution and the distinguishability \citep{Rychtarikovaetal2017b}. In microscopy, the resolution is commonly defined as a distance at which two first-order valleys of the Airy waves~\citep{Airy1848} exactly merge. In a more precise way, the term resolution related to the solution of the inverse problem, i.e., to a function which gives the mechanically determined object's shape of the object from the diffraction pattern~\citep{Resolutionbook}. However, in a real microscope, the Airy pattern is not formed, even for the simplest cases such as the bead in Fig.~3. Even the most advanced descriptions of the inverse problem still utilize the concept of a single focal plane~\citep{Resolutionbook} or several simplifications~\citep{BraatJannsen2015}. Nevertheless, irrespective of the theory, the positions of the electromagnetic centroids of particle's response can be always found. In other words, searching for electromagnetic centroids is a realistic approach in micrographs processing, while the theoretical concept of resolution is misleading and obscures not only the interpretation of the image but even the microscope constructions. 


The interpretation of the fluorescence image is much simpler. 
We search for the distribution of this pointwise information in a cell. In our experiments, the example of the introduced fluorescence is the DAPI labelling. Fig.~7 shows pixels whose intensity was not changed between two z-levels. For the DAPI labelling (Fig.~7\textbf{c}), the objects were selected upon following assumptions:
\begin{enumerate} 
\item Each voxel brings information about presence of fluorophore which is expressed by the image intensity level which, in addition, represents fluorophore's properties.
\item If we observe signals in neighbouring pixels, these pixels can be grouped into an object. Isolated signals, despite having the same intensity as the majority, are assigned to the unspecific labelling and thus to the background. 
\item It was found that there are (i) high intensity points in the centre of the dense areas which most likely represent several fluorophores per voxel which we assigned to the nuclei and (ii) voxels of different intensities, mostly higher, outside dense regions, which we assigned to obviously specific labelling to a DNA-containing object outside the nuclear structures. (iii) Specific labelling which gives rise to the average intensity cannot be distinguished from the unspecific labelling as it does not carry the information which would be distinguishable by a given experiment. 
\end{enumerate}

When an image comprises many fluorescent objects of different intensities, which is typical for autofluorescence (Fig.~7\textbf{a--b}), we cannot use any common simple rule of data analysis. In this case, the location of the chromophore and 3D analysis of the observed structures is possible by the selection of objects of zero point divergence gain. 

The method of electromagnetic centroids determination provide the objective extent of the focal region (not only a single plane) and the proper objective alignment of images overcome the problem. The determination of the in-focus superresolved -- i.e., down to pixel-sized -- objects at each z-level (e.g., Figs.~7--8) further informs us about the co-localization and provides a firm ground for addressing biological questions. We believe that this type of analysis is essential for the proper interpretation of biological data. Moreover, standard widefield microscopy may be used, the chromophores are not bleached, samples are preserved and the analysis may be repeated.

\section{Conclusion}
\label{sec:conclusions}

This article discusses the substance of the solution of the inverse problem in real systems, in contrast to theoretical analyses which dominate the contemporary scientific literature~\citep{Resolutionbook}. We introduce the term electromagnetic centroid for the most localized representation of the observed object in the electromagnetic field interference pattern detected by a digital camera. We introduce the main mathematical and software tools for extraction of the electromagnetic centroids and correction of non-idealities in optical path.

We have also stressed some technical aspects, which are unknown to common microscopists and whose corrections enable the complete yield of information from a microscopic image. The main findings are:
\begin{enumerate} 
\item Usage of a 12-bit-per-channel colour camera equipped with commercial control software may not be sufficient, a higher-bit depth may be necessary. But to work with standard cameras providing 8-bit images is not correct in any known existing set-ups.
\item Usage of a high light intensities leads to the suppression of the noise which makes the extraction of information on localization of an object easier. 
\item There is no theoretical limit for the localization of information. The microscope requires digital cameras with a high number of pixels and small size of the theoretical projected pixel at a higher magnification. This problem was not encountered when photographic films were used for image recording and was not recognized in the early days of the theoretical analyses of microscopic resolution. A low number of pixels on the camera can lead to misinterpretations in many aspects.
\item Not only one in-focus image, but the range of images with all detected in-focus objects has to be determined for each colour channel (i.e. for each wavelength) separately. For proper analysis, these focal regions have to be aligned. Analysis of a dataset not only in planar projection but in the spatial distribution provides near-complete information about specificity and co-localization of fluorescent labels and, eventually, an ultimately correct interpretation. To obtain correct results in this field, we suggest to apply an algorithm based on an information-entropic approach, namely the calculation of the point divergence gain and relevant entropies which enables us to fully describe the image multifractality and extract the information on the response of each observed object.
\end{enumerate}

Samples in biology are usually expensive and often irreplaceable. It is a rather bad idea to image them in a way by which the maximum information from an optical microscopic experiment cannot be acquired and analyzed. The breakdown provided in this paper gives a recipe for how to collect the maximum information and how to interpret it with a negligible loss of intelligibility. We illustrate that results comparable to those obtained by the most advanced microscopes may be obtained from a simple set-up, the widefield microscopy, which is significantly milder to the sample and does not cause photodamage or bleaching.

\section*{Conflict of Interest Statement}
The authors declare that the research was conducted in the absence of any commercial or financial relationships that could be construed as a potential conflict of interest. Author Georg Steiner was employed by company TissueGnostics GmbH, Vienna, Austria. All other authors declare no competing interests.

\section*{Author Contributions}

RR developed the image-processing algorithm, processed the data, and significantly participated in preparation of the manuscript. GK provided samples of cancer tissue sections. GS provided image data of fluorescently labelled tissue. MBF networked the co-authors. D\v{S} is an intellectual co-author of the image processing algorithm and wrote the first version of the manuscript which was further edited by other co-authors.

\section*{Ethics approval and consent to participate}
Cancer tissue sections were provided in agreement with ethic rules of the Medical University Vienna and the study was performed according to guidelines of Good medical practice.

\section*{Funding}
This work was supported by the Ministry of Education, Youth and Sports of the Czech Republic---projects CENAKVA (No. CZ.1.05/2.1.00/01.0024), CENAKVA II (No. LO1205 under the {NPU} I program), the CENAKVA Centre Development (No. CZ.1.05/2.1.00/19.0380)---and from the European Regional Development Fund in frame of the project Kompetenzzentrum MechanoBiologie (ATCZ133) in the Interreg V-A Austria--Czech Republic programme and by TA \v{C}R Gama TG 03010027 subproject 03-24.

\section*{Acknowledgments}
In this paper, the authors used the same image sets as published in~\citep{Rychtarikovaetal2017b} and presented at the 5th International Work-Conference on Bioinformatics and Biomedical Engineering (IWBBIO 2017). The authors would like to thank Jacqueline Dragon for editing and proofreading the manuscript and to Pavel Sou\v{c}ek and Petr Mach\'{a}\v{c}ek for development of the VerCa$\_$cmd software.

\section*{Supplemental Data}
\textbf{Video S1.} The run through a z-stack of $\Omega_{\alpha,k/l}$ = 0 selected (\textit{number-coded}) from the 3D stack of the section of prostate cancer tissue. Upper left corner of Fig.~4. Colour coding of localized and co-localized fluorescent labels: red -- red autofluorescence, green -- green autofluorescence, blue -- DAPI, magenta -- DAPI + red autofluorescence, yellow -- red autofluorescence + green autofluorescence, cyan -- DAPI + green autofluorescence, black -- all three colour channels. The imaging of individual colours was constrained to the regions in which in-focus points were identified by the PDGE analysis ($\alpha = 6$ for red and green autofluorescence and $\alpha = 7$ for DAPI).\\
\textbf{Video S2.} The run through a z-stack of $\Omega_{\alpha,k/l}$ = 0 selected (\textit{number-coded}) from the 3D stack of the section of prostate cancer tissue. Lower left corner of Fig.~4. Colour coding of localized and co-localized fluorescent labels: red -- red autofluorescence, green -- green autofluorescence, blue -- DAPI, magenta -- DAPI + red autofluorescence, yellow -- red autofluorescence + green autofluorescence, cyan -- DAPI + green autofluorescence, black -- all three colour channels. The imaging of individual colours was constrained to the regions in which in-focus points were identified by the PDGE analysis ($\alpha = 6$ for red and green autofluorescence and $\alpha = 7$ for DAPI).

\section*{Data Availability Statement}
The original and processed image sets, relevant Matlab algorithms, and the VerCa$\_$cmd.exe software are available at ftp://160.217.215.251:21/NewMicroscopy (user: anonymous; password: anonymous) or ftps://160.217.215.193:13332/data/NewMicroscopy (user: anonymous; password: anonymous).

\section*{References}
\bibliographystyle{MandM}
\bibliography{references}

\newpage

\begin{figure}
\centering
\includegraphics[width=0.7\textwidth]{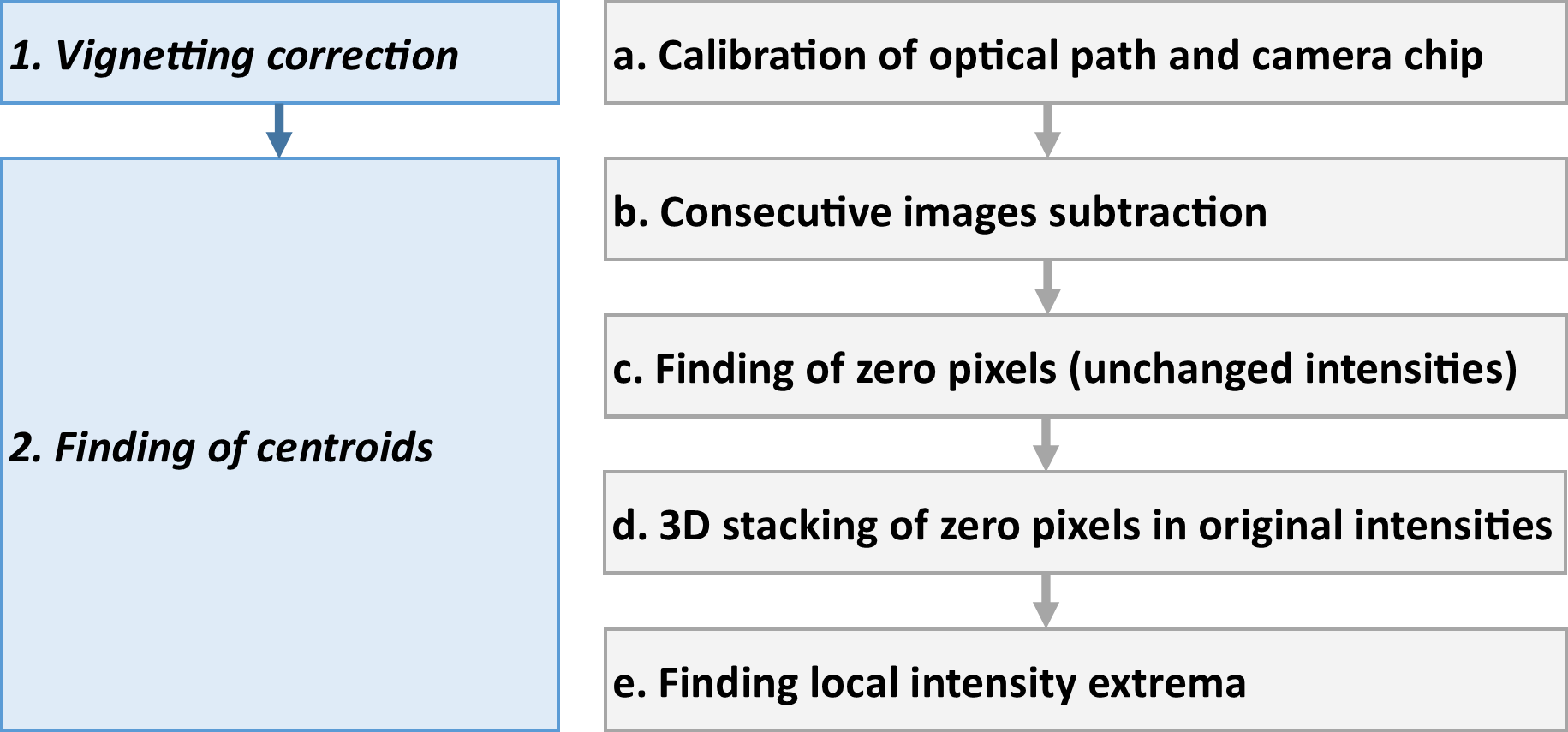}
\caption{Processing of the bright-field microscopic z-stack of the experiment on the 2-$\mu$m bead. The gray boxes on the right state the mathematical procedures leading to the image processing steps written in the left blue boxes.} 
\label{Fig1}
\end{figure}

\begin{figure}
\centering
\includegraphics[width=0.7\textwidth]{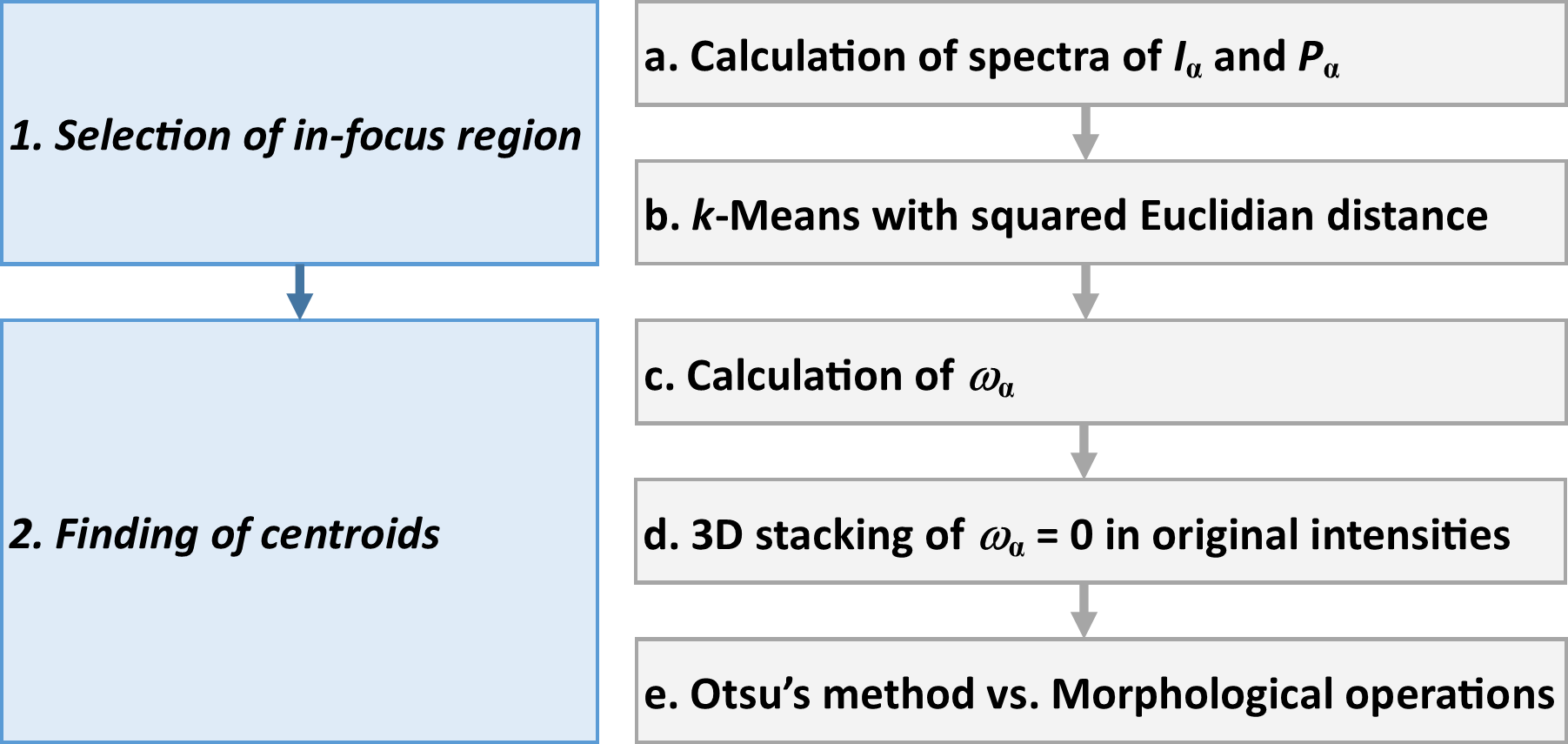}
\caption{Processing of the fluorescence microscopic z-stack of the experiment on the tissue section. The gray boxes on the right state the mathematical procedures leading to the image processing steps written in the left blue boxes.} 
\label{Fig2}
\end{figure}

\begin{figure}
\centering
\includegraphics[width=0.7\textwidth]{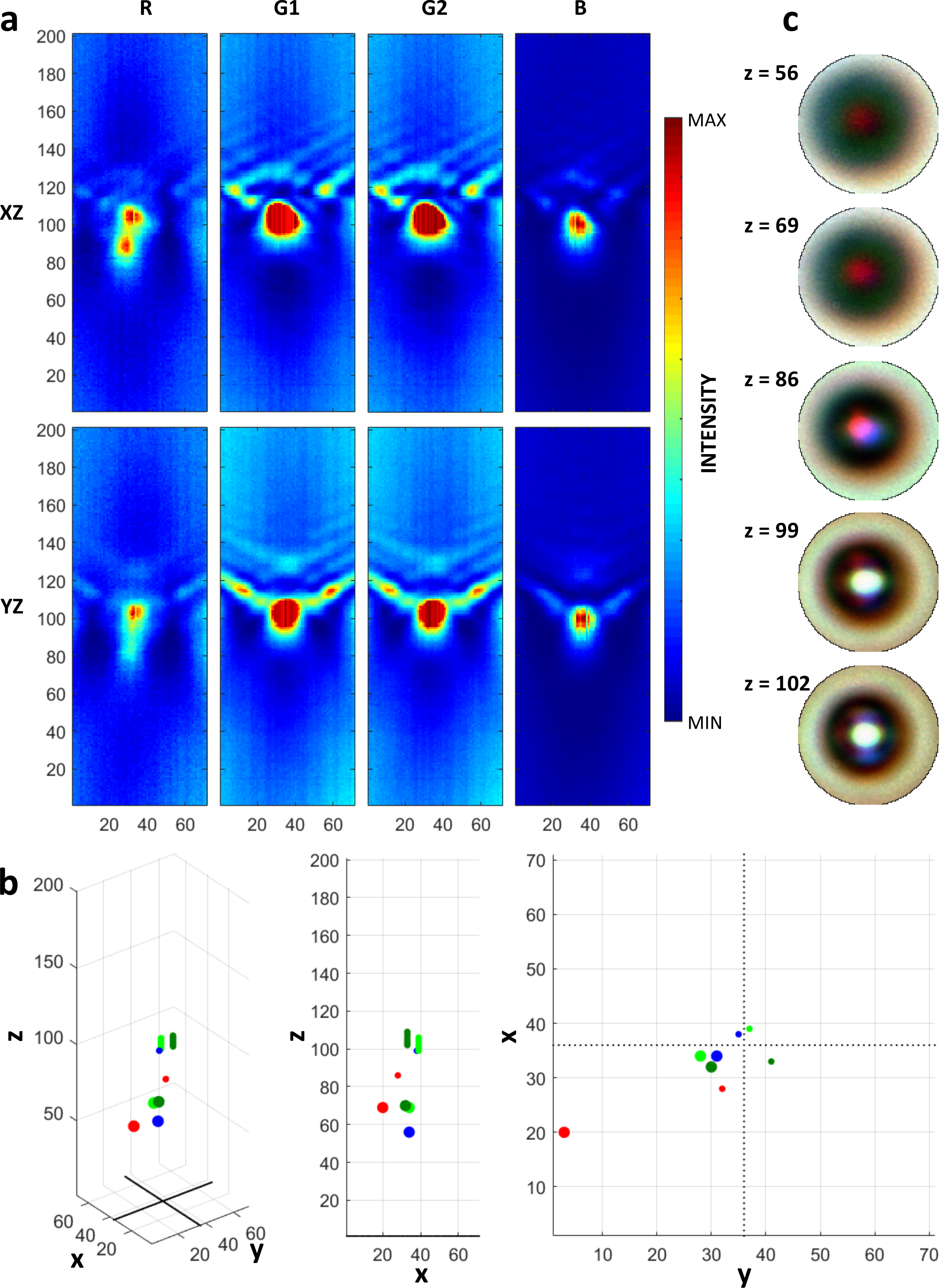}
\caption{3D intensity maps of the 2000-nm latex bead obtained from the red (R), green (G1, G2), and blue (B) pixels of the camera Bayer mask. The data are visualized in a quarter resolution compared with the original data. The minimal/maximal intensities are 89/2982 (R), 371/5710 (G1), 384/5465 (G2), and 19/5908 (B). Voxel size is 46 nm (horizontally) and 152 nm (vertically). \textbf{a)} Sections of the 3D intensity map in the xz- and yz-plane, respectively. \textbf{b)} Positions of electromagnetic centroids (\textit{colour-coded}) in xyz-space (\textit{left}), xz-plane (\textit{middle}), and yx-plane (\textit{right}). The bigger points and the smaller points correspond to negative and positive light interferences, respectively. The positions of xz- and yz-planes relevant to \textbf{a} are highlighted by dotted black bottom lines. \textbf{c)} The images from a z-stack of microscopic images of a 2000-nm latex particle where electromagnetic centroids (see \textbf{b}) were depicted.}
\label{Fig3}
\end{figure}

\begin{figure}
\centering
\includegraphics[width=0.8\textwidth]{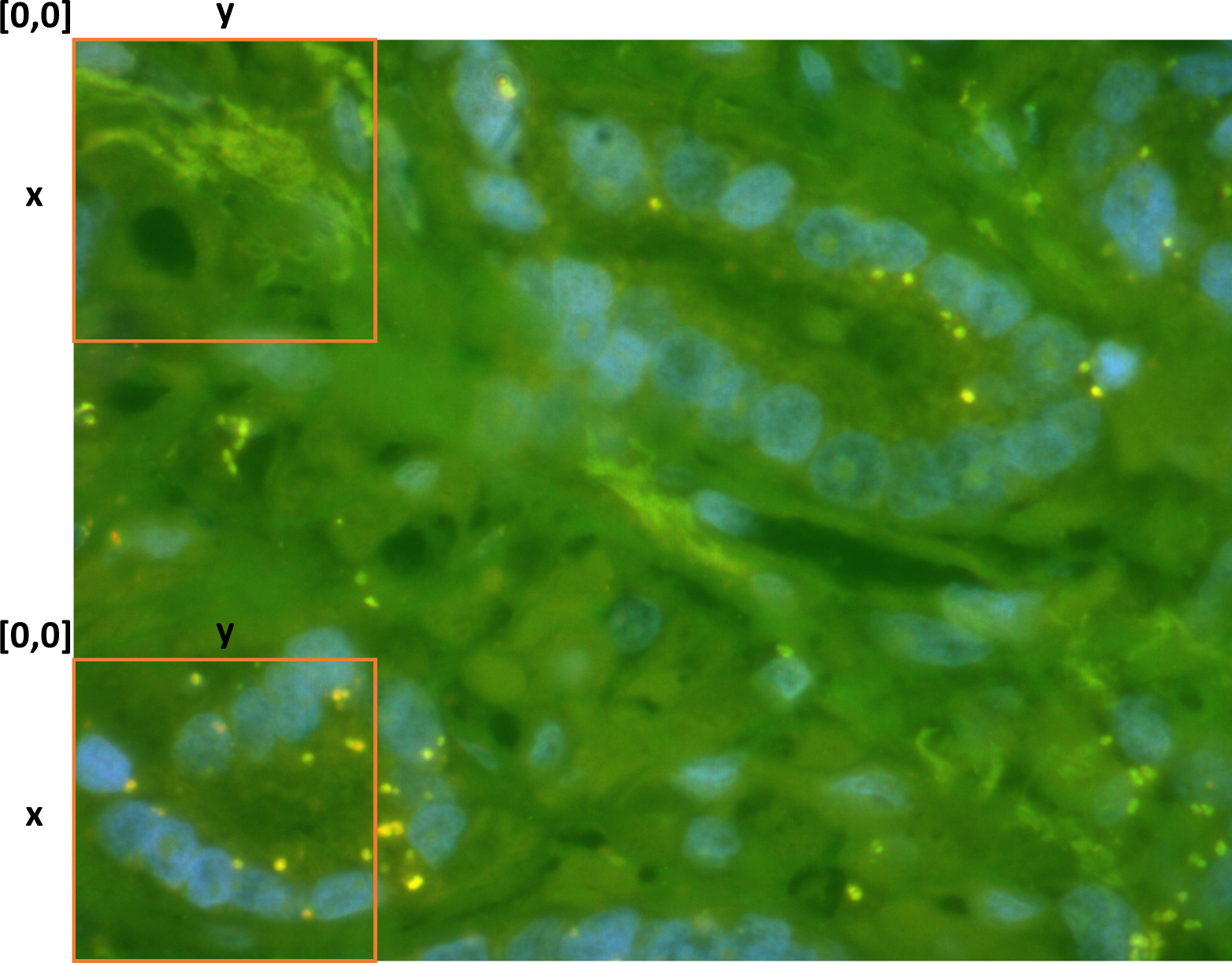}
\caption{Microscopic image of the section of prostate cancer tissue autofluorescing in red and green region of visible spectrum and of DAPI targeted to nuclei. Pixel size is 328$\times$328 nm$^2$. The image was obtained by standard autofocusing. The regions selected by orange squares are those which are further depicted in Figs.~6--8.} 
\label{Fig4}
\end{figure}

\begin{figure}
\centering
\includegraphics[width=0.95\textwidth]{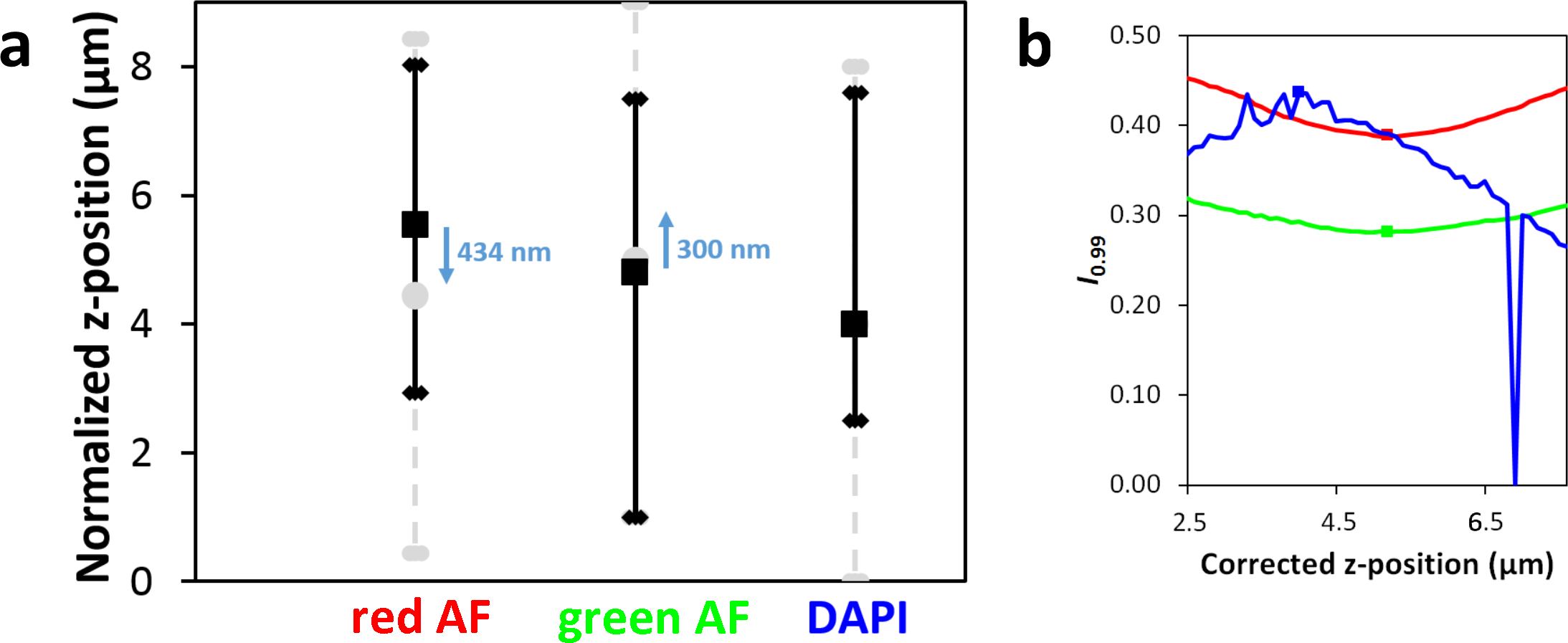}
\caption{Procedure of colocalization of red and green autofluorescence with DAPI in prostate tissue section based on the information-entropic analysis. The real z-positions are normalized on the beginning of the original z-stack of the DAPI-labelled images (zero z-position). \textbf{a)} Ranges of the scanned z-stacks (\textit{gray}) and subseries (focal regions) which were acquired by clustering of the $\alpha$-dependent $I_{\alpha}$ and $P_{\alpha}$ spectra (\textit{black}). The positions of the standard and information foci are marked by gray circles and black squares, respectively. The blue arrows depict the direction and size of the shift of the focal regions to obtain the 3D co-localized maps. \textbf{b)} Paths of the co-localized (corrected) focal regions characterized by the $I_{0.99}$.}
\label{Fig5}
\end{figure}

\begin{figure}
\centering
\caption{The pixels of zero $\Omega_{6.0}$ demonstrating in-focus red autofluorescing objects. Optical cuts (\textit{number-coded}) of the series from which Fig.~7\textbf{a} was constructed. Pixel size is 328$\times$328 nm$^2$.}
\label{Fig6}
\end{figure}

\begin{figure}
\centering
\includegraphics[width=0.7\textwidth]{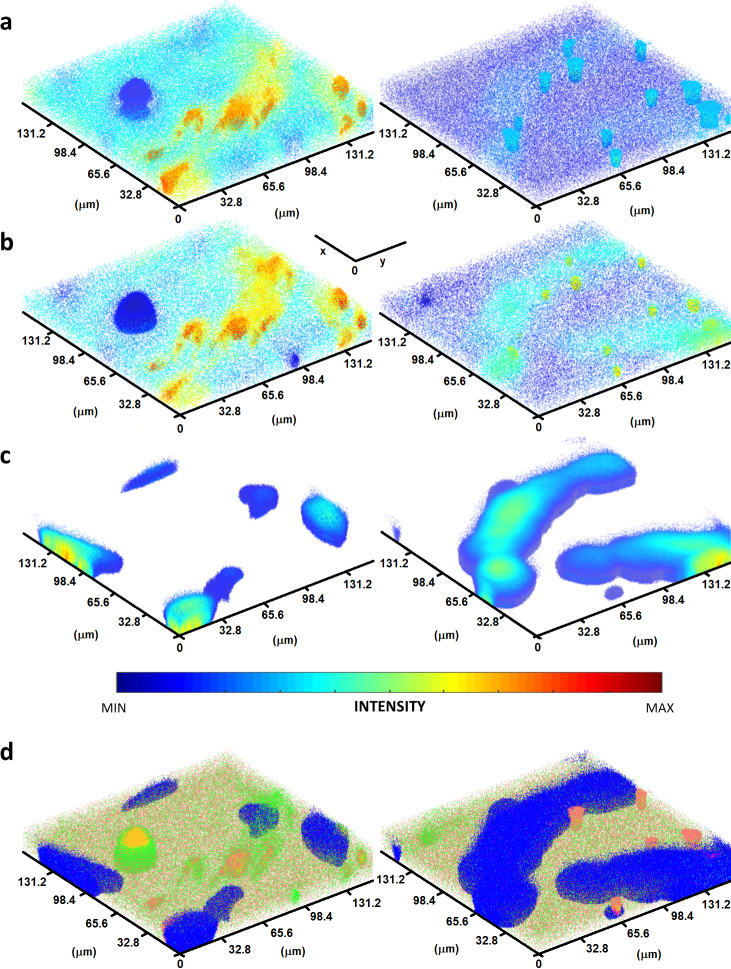}
\caption{3D reconstruction of the positions of points of $\Omega_{\alpha,k/l}$ = 0 in the section of prostate cancer tissue \textbf{a)} autofluorescing in red ($\alpha = 6$), \textbf{b)} autofluorescing in green ($\alpha = 6$), and \textbf{c)} DAPI targeted to nuclei ($\alpha = 7$) and \textbf{d)} the 3D co-localization of all three fluorescent labels. Reconstructions from upper left (\textit{left}) and lower left (\textit{right}) corner of images (see Fig.~6). Colorbars in \textbf{a}--\textbf{c} are rescaled in the range of (the minimal/maximal) intensities of 248/1522, 1689/5378, and 684/3495 for the red autofluorescence, green autofluorescence, and DAPI in the \textit{left} column, respectively, and of 472/2665, 1924/6505, and 661/5969 for red autofluorescence, green autofluorescence, and DAPI in the \textit{right} column, respectively. The pixels of the (co)localization of red autofluorescence, green autofluorescence, DAPI, red autofluorescence + green autofluorescence, red autofluorescence + DAPI, green autofluorescence + DAPI, and red autofluorescence + green autofluorescence + DAPI in \textbf{d} are dyed by red, green, blue, yellow, magenta, cyan, and gray, respectively. The imaging of individual colours was constrained to the regions in which in-focus points were identified by the PDGE analysis.} 
\label{Fig7}
\end{figure}

\begin{figure}
\centering
\includegraphics[width=0.8\textwidth]{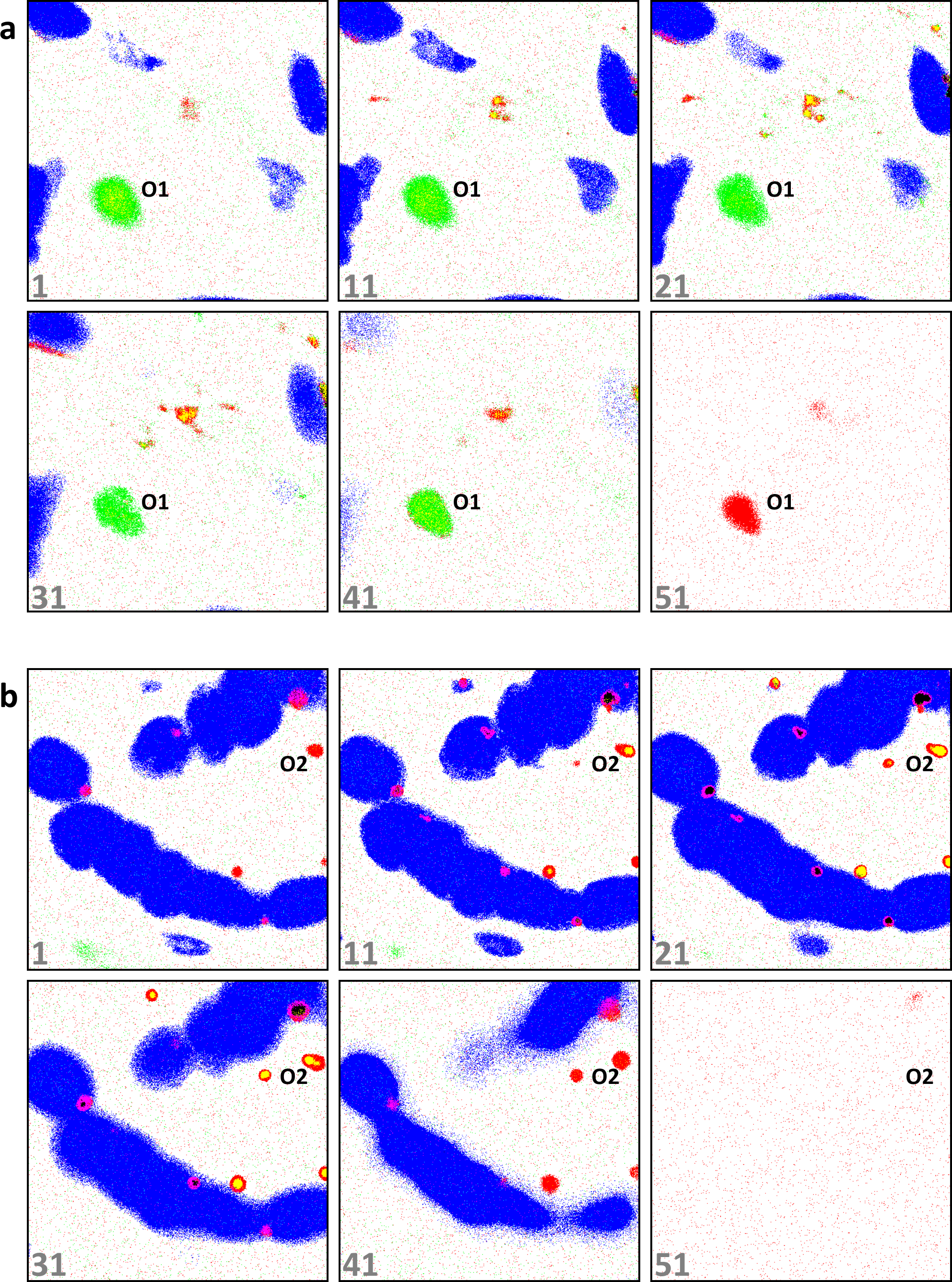}
\caption{Images of $\Omega_{\alpha,k/l}$ = 0 selected (\textit{number-coded}) from the 3D stack of the section of prostate cancer tissue. \textbf{a)} Upper left and \textbf{b)} lower left corner of Fig.~6. Colour coding of localized and co-localized fluorescent labels: red -- red autofluorescence, green -- green autofluorescence, blue -- DAPI, magenta -- DAPI + red autofluorescence, yellow -- red autofluorescence + green autofluorescence, cyan -- DAPI + green autofluorescence, black -- all three colour channels. The imaging of individual colours was constrained to the regions in which in-focus points were identified by the PDGE analysis ($\alpha = 6$ for red and green autofluorescence and $\alpha = 7$ for DAPI).} 
\label{Fig8}
\end{figure}






\end{document}